\documentclass[sigconf,screen,final]{acmart}
\AtBeginDocument{%
  \providecommand\BibTeX{{%
    \normalfont B\kern-0.5em{\scshape i\kern-0.25em b}\kern-0.8em\TeX}}}

\setcopyright{acmlicensed}
\copyrightyear{2023}
\acmYear{2023}
\acmDOI{XXXXXXX.XXXXXXX}

\acmConference[]{}
\acmBooktitle{} 
\usepackage{multirow}
\usepackage{array}

\usepackage{graphicx}

\begin{document}

\title{Evaluating the Effectiveness of GPT-4 Turbo in Creating Defeaters for Assurance Cases} 

\author{Kimya Khakzad Shahandashti}
\email{kimya@yorku,ca}

\affiliation{%
  \institution{York University}
  \city{Toronto}
  \country{Canada}
}

\author{Mithila Sivakumar}
\email{msivakum@yorku.ca}
\affiliation{%
  \institution{York University}
  \city{Toronto}
  \country{Canada}
}

\author{Mohammad Mahdi Mohajer}
\email{mmm98@yorku.ca}

\affiliation{%
  \institution{York University}
  \city{Toronto}
  \country{Canada}
}

\author{Alvine B. Belle}
\email{alvine.belle@lassonde.yorku.ca}

\affiliation{%
  \institution{York University}
  \city{Toronto}
  \country{Canada}
}

\author{Song Wang}
\email{wangsong@yorku.ca}

\affiliation{%
  \institution{York University}
  \city{Toronto}
  \country{Canada}
}

\author{Timothy C. Lethbridge}
\email{timothy.lethbridge@uottawa.ca}

\affiliation{%
  \institution{University of Ottawa}
  \city{Ottawa}
  \country{Canada}
}








\renewcommand{\shortauthors}{Khakzad et al.}

\begin{abstract}
Assurance cases (ACs) are structured arguments that allow verifying the correct implementation of the created systems’ non-functional requirements (e.g., safety, security). This allows for preventing system failure. The latter may result in catastrophic outcomes (e.g., loss of lives). ACs support the certification of systems in compliance with industrial standards e.g. DO-178C and ISO 26262. Identifying defeaters —arguments that challenge these ACs — is crucial for enhancing ACs' robustness and confidence. To automatically support that task, we propose a novel approach that explores the potential of GPT-4 Turbo, an advanced Large Language Model (LLM) developed by OpenAI, in identifying defeaters within ACs formalized using the Eliminative Argumentation (EA) notation. 
Our preliminary evaluation assesses the model's ability to comprehend and generate arguments in this context and the results show that GPT-4 turbo is very proficient in EA notation and can generate different types of defeaters. 
\end{abstract}




\maketitle

\section{Introduction}
An assurance case (AC) is a structured hierarchy of claims aiming at demonstrating that a given mission-critical system supports specific requirements (e.g., safety, security, and privacy)~\cite{hawkins2015weaving, cioroaica2022toward, belle2023evidence}. ACs can be presented in various formats, such as straightforward text like structured prose or through graphical representations. 
Graphical notations include GSN (Goal Structuring Notation)~\cite{GSNV3} and EA (Eliminative Argumentation) \cite {goodenough2015eliminative}).
The presence of assurance weakeners in ACs reflects insufficient evidence, knowledge, or gaps
in reasoning~\cite{hawkins2011new}. These weakeners can undermine confidence in assurance arguments, 
which may hamper the verification of mission-critical system capabilities and further result in catastrophic outcomes~\cite{langari2013safety,rushby2013logic,rushby2014mechanized}. 

Khakzad et al. \cite{shahandashti2023prisma} classified these assurance weakeners by considering several categories, e.g., argument, aleatory, epistemic, and ontological uncertainty. Our focus is on argument uncertainty also referred to as defeaters.
 Inaccurate, incomplete, or inherently flawed reasoning regarding evidence can introduce defects known as argument uncertainty into safety arguments  \cite{muram2018preventing}. This may lead to overconfidence in a system and to the tolerance of certain faults, ultimately contributing to safety-related system failure  \cite{muram2018preventing}.
Manually creating and challenging arguments is recognized as being labour-intensive, time-consuming, and prone to errors \cite{menghi2023assurance,nair2014extended}. 

A few approaches (e.g., \cite{groza2015formal,denney2015dynamic,murugesan2023semantic,yuan2016automatically,viger2023supporting}) allow identifying defeaters in ACs. However, they often fall short of an all-encompassing strategy that covers all types of assurance weakeners, highlighting the urgent requirement for a more integrated identification approach. 
To address that gap, we rely on LLMs to automatically generate defeaters in ACs represented using EA. In our work, we adopt GPT-4 Turbo, owing to its increased efficiency in producing responses and its capability to yield deterministic outputs \cite{OpenAI2023DevDay}. 


\section{Background and related work} \label{sec2}
\subsection{Assurance Cases}

An assurance case (AC) is a \textit{``set of auditable claims, arguments, and evidence created to support the claim that a defined system/service can satisfy particular requirements''} \cite{SACM2021}. An AC is crucial for facilitating clear communication among different stakeholders in a system, such as suppliers and acquirers, and between operators and regulators \cite{SACM2021}. Its primary role is to effectively convey information about the system's non-functional requirements (e.g., safety, security, and reliability)~\cite{belle2023evidence,hawkins2015weaving,health2012evidence}. 
Employing an AC to demonstrate the correct implementation of a system's requirements is crucial to prevent system failure. The latter could have severe consequences such as loss of lives and financial losses~\cite{langari2013safety,de2016industrial}. Hence, several industry standards, including DO-178C in avionics \cite{johnson1998178b} and ISO 26262 in the automotive sector, advocate for the use of ACs to support the certification of systems \cite{matsuno2020facilitating, foster2021integration}.  ACs can be presented in various formats, such as straightforward text like structured prose, or through graphical representations \cite{chelouati2023graphical} (e.g, GSN \cite{GSNV3}, CAE  \cite{bishop2000methodology}, and EA \cite {goodenough2015eliminative}).

\subsection{Eliminative Argumentation}

The EA notation allows constructing arguments and evaluating confidence in these arguments by relying on the notion of \textit{defeasible reasoning}~\cite{diemert2020eliminative,goodenough2015eliminative}. 
The latter supports the recursive challenging of claims to progressively eliminate the doubts (defeaters) they may embed and, consequently, increase the confidence in these arguments \cite{diemert2020eliminative}.


An eliminative argument is comprised of five key components, i.e., Claims, Evidence, Inference Rules, Defeaters, and Argument Terminators~\cite{goodenough2015eliminative}. Claims (C) are assertions that require further argumentative support to establish their credibility. Evidence (E) pertains to observations, data, or artifacts that bolster claims. Strategies (S), which outline the method for arranging a group of claims or defeaters, adopt a "top-down" perspective, encapsulating a comprehensive method for substantiating a claim. Inference Rules (IR) are guidelines for logically amalgamating multiple claims or defeaters to back a higher-level claim. An optional "context" element can be included to provide more details about a primary element. Defeaters challenge the credibility of claims, evidence, and inference rules. 

There are three types of defeaters, classified based on the elements they target, i.e.,  \textbf{rebutting defeaters (R)} that offer reasons why a claim might be false, \textbf{undermining defeaters (UM)} that present arguments why evidence might be unreliable, and \textbf{undercutting defeaters (UC)} that pinpoint flaws in an inference rule such that the validity of its premises doesn't necessarily guarantee the truth of its conclusion. 

\begin{figure}[t!]

\includegraphics[width=\columnwidth]{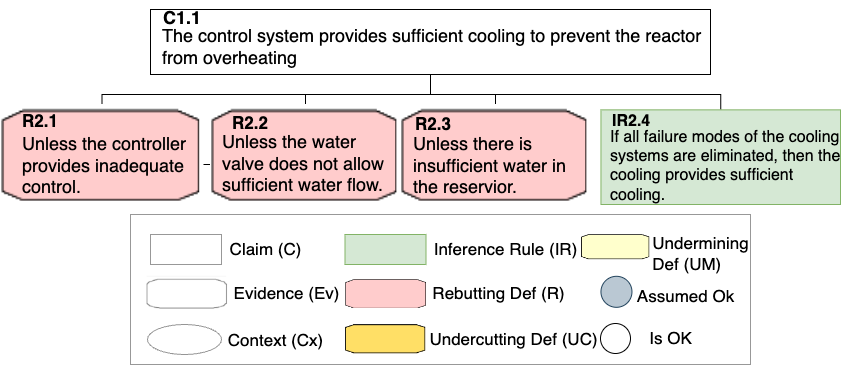}
  \caption{Fragment of EA assurance case adapted from \cite{diemert2020eliminative}}
\label{fig:EASAMPLE}
\end{figure}

Regarding Argument Terminators, the ``Assumed OK'' terminator signifies that further argument or evidence is unnecessary for a defeater, as its resolution is considered obvious. Conversely, the ``Is OK'' terminator is used for an inference rule, indicating it's a tautology without undercutting defeaters, where the premise is deductively equivalent to the conclusion. Figure~\ref{fig:EASAMPLE} 
provides a fragment of an AC in EA notation for a chemical reactor.

\subsection{LLMs and Their Applications}

Large language models (LLMs) are advanced AI models that have become prominent in natural language processing (NLP). Typically built as transformer models, like GPT-4, they are trained on extensive datasets, enabling them to generate text and respond to queries with notable accuracy. Key examples of LLMs include the GPT series by OpenAI, such as GPT-3.5 and GPT-4 \cite{GPT4}, and Google's BERT \cite{devlin2018bert} (Bidirectional Encoder Representations from Transformers). The focus on GPT, particularly GPT-4 Turbo, stems from its advanced capabilities and widespread application potential. GPT-4 Turbo is an enhanced version of the GPT-4 model, known for its larger number of parameters and improved efficiency in generating responses. It works by processing input text and generating responses based on its training, employing a vast number of parameters. GPT-4 Turbo is not inherently deterministic, meaning its responses can vary even with identical prompts. To achieve more deterministic outcomes, techniques like fixing the seed in the random number generator or employing consistent prompting strategies can be used. 


Recent studies have demonstrated various applications of Large Language Models to automate Software Engineering tasks. Chen et al. \cite{chen2023use} focused on GPT-4's use in requirements engineering, specifically for generating goal-oriented models in compliance with the Goal-oriented Requirement Language (GRL). Their work highlighted GPT-4's substantial understanding of goal modeling. Chaaben et al. \cite{chaaben2023towards} utilized ChatGPT to generate UML models, introducing a novel method that employs few-shot prompt learning, thereby reducing the need for large datasets in domain modeling. Viger et al. \cite{viger2023supporting} proposed using GPT-4 to identify defeaters in ACs to enhance their reliability. However, their research is still in its early stages and has not been empirically validated yet. Mahdi Mohajer et al. \cite{mohajer2023skipanalyzer} introduced SkipAnalyzer which is a tool that leverages an LLM-powered agent for static code analysis. It autonomously detects and patches bugs filters out false positives, and is built on ChatGPT. Lastly, Sivakumar et al. \cite{sivakumar2023ArXiv} adapts the work of Chen et al. \cite{chen2023use} to investigate the generation of safety cases using GPT-4, focusing specifically on its understanding of GSN.



\section{Approach} 
\label{sec3}

\begin{figure}[t!]
  \centering
\includegraphics[width=0.786\columnwidth]{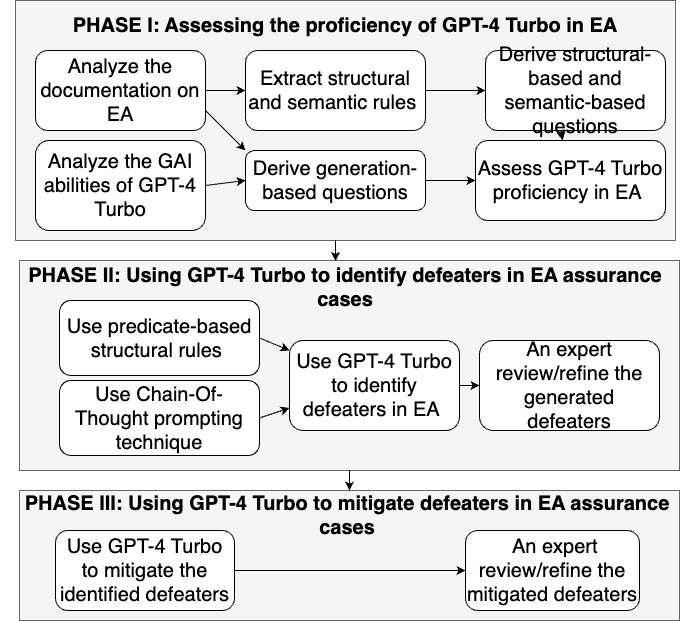}
  \caption{Overview of the Approach}
\label{fig:approach_overview}
\end{figure}

Our work adapts the one of Chen et al. \cite{chen2023use} and Sivakumar et al. \cite{sivakumar2023ArXiv} to the context of ACs formalized with EA. By using an LLM (i.e. GPT-4 Turbo) to automatically identify and mitigate defeaters in ACs, our objective is to emulate some argumentative and doubt-driven aspects of EA \cite{diemert2020eliminative} to better support requirements
verification and validation. Figure \ref{fig:approach_overview} shows a high-level overview of our proposed approach that leverages GPT-4 Turbo to generate (identify) and mitigate defeaters for ACs represented using EA. 
As shown in this figure, our approach consists of three phases. In this paper, we focus on Phase I. Future work will focus on Phase II and Phase III.

In \textbf{Phase I}, like Sivakumar et al. \cite{sivakumar2023ArXiv}, we conduct a thorough analysis of the documentation on EA (e.g., \cite{goodenough2015eliminative, diemert2020eliminative}) to extract the structural and semantic rules its notation embodies. We then derive structural and semantic-based questions from these rules. We combine these questions with EA generation-based questions. We use the resulting set of questions to assess GPT-4 Turbo's proficiency in EA by challenging its understanding of the syntax and semantics of EA, as well as its ability to generate EA concepts.

\textbf{Phase II} is centered around applying GPT-4 Turbo to identify potential defeaters within EA assurance cases. We plan to guide GPT-4 Turbo through Chain-Of-Thought prompting techniques to clarify the reasoning behind defeater identification. This involves either integrating the reasoning steps into the examples or soliciting a detailed explanation of the thought process from the LLM \cite{wei2022chain, kojima2022large}. Moreover, we plan to incorporate a strategy similar to that of Zhu, Zhaocheng, et al. \cite{zhu2023large}, who proposed the Hypotheses-to-Theories (HtT) framework to establish a rule library for structured reasoning with LLMs. This is because prompting methods relying on an LLM's implicit knowledge are prone to "hallucinations" \cite{mcintosh2023culturally}, leading to incorrect answers. We plan to adopt this technique, forming predicate-based rules from the structural rules EA embodies to prompt GPT-4 Turbo such that it could generate better defeaters. Additionally, we involve an expert to review and refine the defeaters generated by GPT-4 Turbo to ensure their validity and applicability.


\textbf{Phase III} relies on GPT-4 Turbo to mitigate the defeaters identified in Phase II. Like Phase II, an expert is engaged to review/ refine the mitigation strategies proposed by GPT-4 Turbo, enhancing the reliability of ACs and ensuring they withstand rigorous scrutiny.

\section{Experimental setup} \label{sec4}
This preliminary study focuses on Phase I and is designed to assess the proficiency of GPT-4 Turbo in understanding and applying EA. 

\subsection{Research Objective}
The goal of our preliminary study is to answer this research question (RQ): \textbf{Is GPT-4 Turbo sufficiently proficient in the EA notation?} Like Chen et al. \cite{chen2023use} and Sivakumar et al. \cite{sivakumar2023ArXiv}, to investigate this RQ, we formulated 22 specific questions based on the EA documentation. We rely on them to assess the proficiency of GPT-4 Turbo in understanding and applying the structural and semantic rules of EA and its ability to generate EA concepts (e.g., defeaters). 


\subsection{Extraction Of The Structural and Semantic Rules Of EA}

The first step of our approach consists in analyzing existing documentation (e.g., \cite{goodenough2015eliminative, diemert2020eliminative}) on EA to extract the set of structural and semantic rules that EA embodies. Table \ref{tab:rules} reports the structural and semantic rules that we extracted. Extracting rules is a critical component of our research, forming the core foundation for crafting essential questions relevant to RQ. Semantic rules are focused on the text within EA elements and the meaning of that text. On the other hand, structural rules address the arrangement and design of EA elements, conveying the appropriate structure of each element and the nature of its relationships with other elements. 

{\small
\begin{table*}[t!]
  \caption{EA Structural and Semantic rules}
  \label{tab:rules}
  \begin{tabular}{p{1.5cm} p{1.5cm} p{5cm} p{7.5cm}}
    \toprule
    \textbf{Category} &  \textbf{Name} & \textbf{Structural rules} & \textbf{Semantic rules} \\
    \midrule
    EA Element & Claim & Connected to: Context, Rebutting Defeater & A claim is stated as a predicate, a true or false statement. \\
   EA Element & Evidence & 
   Connected to: Rebutting defeater, Undermining defeater, Undercutting defeater, Inference rule, Evidence & Evidence is in the form ``[Noun phrase] showing P'' with P asserting an interpretation of data relevant to the argument. \\
    EA Element & Context & Connected to: Claim & It gives additional information about the content of a fundamental element and is optional. \\
    EA Element & Inference Rule & Connected to Rebutting defeater, Undermining defeater, Undercutting defeater, Claim, Evidence & They are predicates (P → Q), where either P or Q (but not both) is an eliminated defeater. \\
    EA Element & Undercutting Defeater & Connected to: Inference Rule & Is a doubt about the validity of an inference rule (P → Q), preceded by ``Unless''\\
    EA Element & Undermining Defeater &  Connected to Evidence &Is a predicate associated with evidence, preceded by "But". It challenges the validity of the data comprising the evidence.\\
    EA Element & Rebutting Defeater & Connected to: Claim & Is a predicate associated with a claim, preceded by ``Unless'' \\
    Argument Terminator & Assumed OK & Attached to Rebutting defeater, Undermining defeater, Undercutting defeater, Claim, Evidence & It asserts that some defeater is (assumed to be) false.\\
    Argument Terminator & Is OK & Attached to Inference Rule, Claim, Evidence & It applies to inference rules, indicating no undercutting defeaters due to the rule being a tautology.\\

    \bottomrule
  \end{tabular}
\end{table*}
}

\subsection{Generation Of Questions To Assess GPT-4 Turbo's Proficiency In EA}

We crafted an initial batch of 22 questions categorized in three sections, i.e., structural, semantic, and generation-based questions. The structural questions aim to assess GPT-4 Turbo's comprehension of EA notation i.e. EA structural rules. 
In contrast, the semantic questions examine its understanding of the semantic rules of EA. Lastly, the generation-focused questions are designed to test GPT-4 Turbo's capability in effectively generating EA elements, with a specific emphasis on defeaters. We developed eight structural, seven semantic, and seven generation-based questions. 
The complete list of these questions together with supplemental material is available on GitHub\footnote{Supplemental material link: \url{https://github.com/kimixz/GPT4-TURBO-EFFICIENCY}}. Table \ref{tab:qns} presents sample questions for each category. 

{\small
\begin{table*}
  \caption{Sample Questions for assessing GPT-4 Turbo's Proficiency in EA}
  \label{tab:qns}
  \begin{tabular}{p{4cm} p{13cm}}
    \toprule
    Category of Question & Sample Question \\
    \midrule
     Structural Question & What are the different types of defeaters in Eliminative Argumentation? \\ [0.25cm]
    Semantic questions &  How should a claim be structured in Eliminative Argumentation? i.e., mention whether it can be in the form of noun-phrase, verb-phrase or predicate. \\ [0.25cm]
    Generation-based Question & Generate me a sample Claim and a Rebutting defeater that defeats it. Show it in structured prose. \\
    \bottomrule
  \end{tabular}
\end{table*}
}

\subsection{GPT-4 Turbo Setting}

In our study, we utilized the OpenAI API\footnote{\url{https://openai.com/api/}} to interact with the GPT-4 Turbo model. A key methodological decision was ensuring deterministic responses from the model. This was achieved by setting the \texttt{seed} parameter in the API. The \texttt{seed} parameter is vital for generating consistent outputs from GPT-4 Turbo for the same input. It initializes the model's internal random number generator to a fixed state, thereby ensuring reproducibility and consistency of the responses generated by GPT-4 Turbo for identical prompts.

\subsection{Prompting Process}

In our study, we followed the best practices of prompt engineering as outlined in OpenAI's guide\footnote{\url{https://platform.openai.com/docs/guides/prompt-engineering}} to interact with GPT-4 Turbo. The process involved careful construction of both 'system' and 'user' prompts to guide the model effectively. The 'user' prompts were the direct questions posed to the model, designed to extract specific information or analysis. Meanwhile, for 'system' prompts, the primary objective was to orient GPT-4 Turbo appropriately, ensuring it understood its role as an assistant in addressing our inquiries. The system prompt that we used is provided in the box below:

\smallskip
\noindent\fbox{%
    \parbox{\linewidth}{%
    \textbf{System Prompt:}
    You are an assistant that helps me answer questions about Eliminative Argumentation.
Eliminative Argumentation is a method used in ACs, particularly in the fields of software engineering and system safety. It focuses on systematically identifying and eliminating potential causes of failure to strengthen the assurance of system safety and reliability.
Answer each question separately and try to generate the samples even if they are simple. Your answers should be concise and to the point. It should not be more than 2-3 lines.
    }}
    


\subsection{Assessment Process}
GPT-4 Turbo generates each EA concept in the structured prose complying with EA. To evaluate each of the GPT-4 Turbo's responses to the 22 questions, we had two researchers with extensive expertise in EA to assess these responses. They independently rated each of GPT-4 Turbo's answers on a linear scale ranging from one (totally correct) to five (incorrect). To assess the consistency and reliability of these ratings, we rely on the Kendall rank correlation coefficient as in Chen et al. \cite{chen2023use}. 
The value of that correlation
coefficient progresses from - 1 to 1.
A value close to -1 means the level of agreement between raters is almost close to none. A value close to 1 means the level of agreement is strong. We rely on an online tool i.e. \textbf{Gigacalculator}\footnote{\href{https://www.gigacalculator.com/calculators/correlation-coefficient-calculator.php}{https://www.gigacalculator.com/calculators/correlation-coefficient-calculator.php}} to automatically assess the values of the Kendall rank coefficient with a confidence level of 95\%.

\section{Preliminary Results} \label{sec5}

Two researchers evaluated the answers of GPT-4 to the 22 questions. 
The resulting correlation between their ratings is \textbf{75\%}. That strong correlation level underscores a robust agreement between the two assessors and a high level of consistency in their ratings. 

Table \ref{tab:avgscores} reports the average ratings of the answers GPT-4 Turbo provided to structural, semantic, and generation-based. 
The average of the ratings achieved by GPT-4 Turbo is \textbf{1.40}, which is close to 1. That value reflects its strong grasp of the essential elements of EA notation. In the context of the grading systems used by many universities, that average of ratings equates to a \textbf{grade of A}.

{\small
\begin{table}
  \caption{Average ratings of questions in RQ}
  \label{tab:avgscores}
  \begin{tabular}{cccc}
    \toprule
   Structural & Semantic & Generation-based & Avg\\
    \midrule
    1.125 & 1.78 & 1.35 & 1.40 \\
    \bottomrule
  \end{tabular}
\end{table}
}
\noindent\fbox{%
    \parbox{\linewidth}{%
GPT-4 Turbo achieved excellent proficiency when answering structural-based questions. It also showed commendable performance when answering generation-based questions, effectively creating EA elements, particularly various types of defeaters. However, GPT-4 Turbo's understanding of the semantics of EA elements was less robust. Thus, it would be beneficial to employ specific prompting techniques to enhance its comprehension of EA semantics, which in turn could lead to the generation of better defeaters.}}

\section{Conclusion} \label{sec6}

Our investigation into the capabilities of GPT-4 Turbo has revealed its excellent proficiency in understanding and applying EA notation. This underscores the model's potential as a valuable tool for future endeavors in the identification (Phase II) and mitigation of defeaters (Phase III) within ACs represented using EA. We are optimistic about the role of GPT-4 Turbo in enhancing the robustness of ACs, particularly in mission-critical systems where the assurance of non-functional requirements is paramount. 

\bibliographystyle{ACM-Reference-Format}
\bibliography{sample-authordraft}


\begin{thebibliography}{38}


\ifx \showCODEN    \undefined \def \showCODEN     #1{\unskip}     \fi
\ifx \showDOI      \undefined \def \showDOI       #1{#1}\fi
\ifx \showISBNx    \undefined \def \showISBNx     #1{\unskip}     \fi
\ifx \showISBNxiii \undefined \def \showISBNxiii  #1{\unskip}     \fi
\ifx \showISSN     \undefined \def \showISSN      #1{\unskip}     \fi
\ifx \showLCCN     \undefined \def \showLCCN      #1{\unskip}     \fi
\ifx \shownote     \undefined \def \shownote      #1{#1}          \fi
\ifx \showarticletitle \undefined \def \showarticletitle #1{#1}   \fi
\ifx \showURL      \undefined \def \showURL       {\relax}        \fi
\providecommand\bibfield[2]{#2}
\providecommand\bibinfo[2]{#2}
\providecommand\natexlab[1]{#1}
\providecommand\showeprint[2][]{arXiv:#2}

\bibitem[B.~Belle and Zhao(2023)]%
        {belle2023evidence}
\bibfield{author}{\bibinfo{person}{A. B.~Belle} {and} \bibinfo{person}{Y. Zhao}.} \bibinfo{year}{2023}\natexlab{}.
\newblock \showarticletitle{Evidence-based decision-making: On the use of systematicity cases to check the compliance of reviews with reporting guidelines such as PRISMA 2020}.
\newblock \bibinfo{journal}{\emph{Expert Systems with Applications}}  \bibinfo{volume}{217} (\bibinfo{year}{2023}), \bibinfo{pages}{119569}.
\newblock


\bibitem[Bishop and Bloomfield(2000)]%
        {bishop2000methodology}
\bibfield{author}{\bibinfo{person}{P. Bishop} {and} \bibinfo{person}{R. Bloomfield}.} \bibinfo{year}{2000}\natexlab{}.
\newblock \showarticletitle{A methodology for safety case development}. In \bibinfo{booktitle}{\emph{Safety and Reliability}}, Vol.~\bibinfo{volume}{20}. Taylor \& Francis, \bibinfo{pages}{34--42}.
\newblock


\bibitem[Chaaben et~al\mbox{.}(2023)]%
        {chaaben2023towards}
\bibfield{author}{\bibinfo{person}{M. Chaaben}, \bibinfo{person}{L. Burgue{\~n}o}, {and} \bibinfo{person}{H. Sahraoui}.} \bibinfo{year}{2023}\natexlab{}.
\newblock \showarticletitle{Towards using few-shot prompt learning for automating model completion}. In \bibinfo{booktitle}{\emph{ICSE-NIER}}. IEEE, \bibinfo{pages}{7--12}.
\newblock


\bibitem[Chelouati et~al\mbox{.}(2023)]%
        {chelouati2023graphical}
\bibfield{author}{\bibinfo{person}{M. Chelouati}, \bibinfo{person}{A. Boussif}, \bibinfo{person}{J. Beugin}, {and} \bibinfo{person}{E. El~Koursi}.} \bibinfo{year}{2023}\natexlab{}.
\newblock \showarticletitle{Graphical safety assurance case using Goal Structuring Notation (GSN)—challenges, opportunities and a framework for autonomous trains}.
\newblock \bibinfo{journal}{\emph{Reliability Engineering \& System Safety}}  \bibinfo{volume}{230} (\bibinfo{year}{2023}), \bibinfo{pages}{108933}.
\newblock


\bibitem[Chen et~al\mbox{.}(2023)]%
        {chen2023use}
\bibfield{author}{\bibinfo{person}{B. Chen}, \bibinfo{person}{K. Chen}, \bibinfo{person}{S. Hassani}, \bibinfo{person}{Y. Yang}, \bibinfo{person}{D. Amyot}, \bibinfo{person}{L. Lessard}, \bibinfo{person}{G. Mussbacher}, \bibinfo{person}{M. Sabetzadeh}, {and} \bibinfo{person}{D. Varr{\'o}}.} \bibinfo{year}{2023}\natexlab{}.
\newblock \showarticletitle{On the use of GPT-4 for creating goal models: an exploratory study}. In \bibinfo{booktitle}{\emph{REW}}. IEEE, \bibinfo{pages}{262--271}.
\newblock


\bibitem[Cioroaica et~al\mbox{.}(2022)]%
        {cioroaica2022toward}
\bibfield{author}{\bibinfo{person}{E. Cioroaica}, \bibinfo{person}{B. Buhnova}, \bibinfo{person}{D. Schneider}, \bibinfo{person}{I. Sorokos}, \bibinfo{person}{T. Kuhn}, {and} \bibinfo{person}{E. Tomur}.} \bibinfo{year}{2022}\natexlab{}.
\newblock \showarticletitle{Towards the Concept of Trust Assurance Case}. In \bibinfo{booktitle}{\emph{TrustCom}}. IEEE, \bibinfo{pages}{1581--1586}.
\newblock


\bibitem[de~La~Vara et~al\mbox{.}(2016)]%
        {de2016industrial}
\bibfield{author}{\bibinfo{person}{J.~L. de La~Vara}, \bibinfo{person}{M. Borg}, \bibinfo{person}{K. Wnuk}, {and} \bibinfo{person}{L. Moonen}.} \bibinfo{year}{2016}\natexlab{}.
\newblock \showarticletitle{An industrial survey of safety evidence change impact analysis practice}.
\newblock \bibinfo{journal}{\emph{TSE}} \bibinfo{volume}{42}, \bibinfo{number}{12} (\bibinfo{year}{2016}), \bibinfo{pages}{1095--1117}.
\newblock


\bibitem[Denney et~al\mbox{.}(2015)]%
        {denney2015dynamic}
\bibfield{author}{\bibinfo{person}{E. Denney}, \bibinfo{person}{G. Pai}, {and} \bibinfo{person}{I. Habli}.} \bibinfo{year}{2015}\natexlab{}.
\newblock \showarticletitle{Dynamic safety cases for through-life safety assurance}. In \bibinfo{booktitle}{\emph{ICSE}}, Vol.~\bibinfo{volume}{2}. IEEE, \bibinfo{pages}{587--590}.
\newblock


\bibitem[Devlin et~al\mbox{.}(2018)]%
        {devlin2018bert}
\bibfield{author}{\bibinfo{person}{J. Devlin}, \bibinfo{person}{M. Chang}, \bibinfo{person}{K. Lee}, {and} \bibinfo{person}{Kristina Toutanova}.} \bibinfo{year}{2018}\natexlab{}.
\newblock \showarticletitle{Bert: Pre-training of deep bidirectional transformers for language understanding}.
\newblock \bibinfo{journal}{\emph{arXiv preprint arXiv:1810.04805}} (\bibinfo{year}{2018}).
\newblock


\bibitem[Diemert and Joyce(2020)]%
        {diemert2020eliminative}
\bibfield{author}{\bibinfo{person}{S. Diemert} {and} \bibinfo{person}{J. Joyce}.} \bibinfo{year}{2020}\natexlab{}.
\newblock \showarticletitle{Eliminative Argumentation for Arguing System Safety-A Practitioner’s Experience}. In \bibinfo{booktitle}{\emph{SysCon}}. IEEE, \bibinfo{pages}{1--7}.
\newblock


\bibitem[Foster et~al\mbox{.}(2021)]%
        {foster2021integration}
\bibfield{author}{\bibinfo{person}{Simon Foster}, \bibinfo{person}{Yakoub Nemouchi}, \bibinfo{person}{Mario Gleirscher}, \bibinfo{person}{Ran Wei}, {and} \bibinfo{person}{Tim Kelly}.} \bibinfo{year}{2021}\natexlab{}.
\newblock \showarticletitle{Integration of formal proof into unified assurance cases with Isabelle/SACM}.
\newblock \bibinfo{journal}{\emph{Formal Aspects of Computing}} \bibinfo{volume}{33}, \bibinfo{number}{6} (\bibinfo{year}{2021}), \bibinfo{pages}{855--884}.
\newblock


\bibitem[Foundation(2012)]%
        {health2012evidence}
\bibfield{author}{\bibinfo{person}{Health Foundation}.} \bibinfo{year}{2012}\natexlab{}.
\newblock \bibinfo{title}{Evidence: Using Safety Cases in Industry and Healthcare}.
\newblock
\newblock


\bibitem[Goodenough et~al\mbox{.}(2015)]%
        {goodenough2015eliminative}
\bibfield{author}{\bibinfo{person}{J.~B. Goodenough}, \bibinfo{person}{C.~B. Weinstock}, {and} \bibinfo{person}{A.~Z. Klein}.} \bibinfo{year}{2015}\natexlab{}.
\newblock \showarticletitle{Eliminative argumentation: A basis for arguing confidence in system properties}.
\newblock \bibinfo{journal}{\emph{SEI, Carnegie Mellon University, Pittsburgh, PA, Tech. Rep. CMU/SEI-2015-TR-005}} (\bibinfo{year}{2015}).
\newblock


\bibitem[Group(2021)]%
        {GSNV3}
\bibfield{author}{\bibinfo{person}{The Assurance Case~Working Group}.} \bibinfo{year}{2021}\natexlab{}.
\newblock \bibinfo{title}{Goal Structuring Notation Standard Version 3}.
\newblock
\newblock
\urldef\tempurl%
\url{https://scsc.uk/r141C:1?t=1}
\showURL{%
\tempurl}


\bibitem[Groza et~al\mbox{.}(2015)]%
        {groza2015formal}
\bibfield{author}{\bibinfo{person}{A. Groza}, \bibinfo{person}{I.~A. Letia}, \bibinfo{person}{A. Goron}, {and} \bibinfo{person}{S. Zaporojan}.} \bibinfo{year}{2015}\natexlab{}.
\newblock \showarticletitle{A formal approach for identifying assurance deficits in unmanned aerial vehicle software}. In \bibinfo{booktitle}{\emph{ICSEng}}. Springer, \bibinfo{pages}{233--239}.
\newblock


\bibitem[Hawkins et~al\mbox{.}(2015)]%
        {hawkins2015weaving}
\bibfield{author}{\bibinfo{person}{R. Hawkins}, \bibinfo{person}{I. Habli}, \bibinfo{person}{D. Kolovos}, \bibinfo{person}{R. Paige}, {and} \bibinfo{person}{T. Kelly}.} \bibinfo{year}{2015}\natexlab{}.
\newblock \showarticletitle{Weaving an assurance case from design: a model-based approach}. In \bibinfo{booktitle}{\emph{HASE}}. IEEE, \bibinfo{pages}{110--117}.
\newblock


\bibitem[Hawkins et~al\mbox{.}(2011)]%
        {hawkins2011new}
\bibfield{author}{\bibinfo{person}{R. Hawkins}, \bibinfo{person}{T. Kelly}, \bibinfo{person}{J. Knight}, {and} \bibinfo{person}{P. Graydon}.} \bibinfo{year}{2011}\natexlab{}.
\newblock \showarticletitle{A new approach to creating clear safety arguments}. In \bibinfo{booktitle}{\emph{SSS}}. Springer, \bibinfo{pages}{3--23}.
\newblock


\bibitem[Johnson et~al\mbox{.}(1998)]%
        {johnson1998178b}
\bibfield{author}{\bibinfo{person}{L.~A. Johnson} {et~al\mbox{.}}} \bibinfo{year}{1998}\natexlab{}.
\newblock \showarticletitle{DO-178B: Software considerations in airborne systems and equipment certification}.
\newblock \bibinfo{journal}{\emph{Crosstalk, October}}  \bibinfo{volume}{199} (\bibinfo{year}{1998}), \bibinfo{pages}{11--20}.
\newblock


\bibitem[Khakzad~S. et~al\mbox{.}(2023)]%
        {shahandashti2023prisma}
\bibfield{author}{\bibinfo{person}{K. Khakzad~S.}, \bibinfo{person}{Alvine~B. Belle}, \bibinfo{person}{T.~C. Lethbridge}, \bibinfo{person}{O. Odu}, {and} \bibinfo{person}{M. Sivakumar}.} \bibinfo{year}{2023}\natexlab{}.
\newblock \showarticletitle{A PRISMA-driven systematic mapping study on system assurance weakeners}.
\newblock \bibinfo{journal}{\emph{arXiv preprint arXiv:2311.08328}} (\bibinfo{year}{2023}).
\newblock


\bibitem[Kojima et~al\mbox{.}(2022)]%
        {kojima2022large}
\bibfield{author}{\bibinfo{person}{T. Kojima}, \bibinfo{person}{S.~S. Gu}, \bibinfo{person}{M. Reid}, \bibinfo{person}{Y. Matsuo}, {and} \bibinfo{person}{Y. Iwasawa}.} \bibinfo{year}{2022}\natexlab{}.
\newblock \showarticletitle{Large language models are zero-shot reasoners}.
\newblock \bibinfo{journal}{\emph{NeuRIPS}}  \bibinfo{volume}{35} (\bibinfo{year}{2022}), \bibinfo{pages}{22199--22213}.
\newblock


\bibitem[Langari and Maibaum(2013)]%
        {langari2013safety}
\bibfield{author}{\bibinfo{person}{Z. Langari} {and} \bibinfo{person}{T. Maibaum}.} \bibinfo{year}{2013}\natexlab{}.
\newblock \showarticletitle{Safety cases: a review of challenges}. In \bibinfo{booktitle}{\emph{ASSURE}}. IEEE, \bibinfo{pages}{1--6}.
\newblock


\bibitem[Matsuno et~al\mbox{.}(2020)]%
        {matsuno2020facilitating}
\bibfield{author}{\bibinfo{person}{Yutaka Matsuno}, \bibinfo{person}{Toshinori Takai}, {and} \bibinfo{person}{Shuichiro Yamamoto}.} \bibinfo{year}{2020}\natexlab{}.
\newblock \showarticletitle{Facilitating use of assurance cases in industries by workshops with an agent-based method}.
\newblock \bibinfo{journal}{\emph{IEICE TRANSACTIONS on Information and Systems}} \bibinfo{volume}{103}, \bibinfo{number}{6} (\bibinfo{year}{2020}), \bibinfo{pages}{1297--1308}.
\newblock


\bibitem[McIntosh et~al\mbox{.}(2023)]%
        {mcintosh2023culturally}
\bibfield{author}{\bibinfo{person}{T.~R. McIntosh}, \bibinfo{person}{T. Liu}, \bibinfo{person}{T. Susnjak}, \bibinfo{person}{P. Watters}, \bibinfo{person}{A. Ng}, {and} \bibinfo{person}{M.~N. Halgamuge}.} \bibinfo{year}{2023}\natexlab{}.
\newblock \showarticletitle{A culturally sensitive test to evaluate nuanced gpt hallucination}.
\newblock \bibinfo{journal}{\emph{TAI}} \bibinfo{volume}{1}, \bibinfo{number}{01} (\bibinfo{year}{2023}), \bibinfo{pages}{1--13}.
\newblock


\bibitem[Menghi et~al\mbox{.}(2023)]%
        {menghi2023assurance}
\bibfield{author}{\bibinfo{person}{C. Menghi}, \bibinfo{person}{T. Viger}, \bibinfo{person}{A. Di~Sandro}, \bibinfo{person}{C. Rees}, \bibinfo{person}{J. Joyce}, {and} \bibinfo{person}{M. Chechik}.} \bibinfo{year}{2023}\natexlab{}.
\newblock \showarticletitle{Assurance case development as data: A manifesto}. In \bibinfo{booktitle}{\emph{ICSE-NIER}}. IEEE, \bibinfo{pages}{135--139}.
\newblock


\bibitem[Mohajer et~al\mbox{.}(2023)]%
        {mohajer2023skipanalyzer}
\bibfield{author}{\bibinfo{person}{M.~M. Mohajer}, \bibinfo{person}{R. Aleithan}, \bibinfo{person}{N.~S. Harzevili}, \bibinfo{person}{M. Wei}, \bibinfo{person}{A.~B. Belle}, \bibinfo{person}{H.~V. Pham}, {and} \bibinfo{person}{S. Wang}.} \bibinfo{year}{2023}\natexlab{}.
\newblock \showarticletitle{SkipAnalyzer: An Embodied Agent for Code Analysis with Large Language Models}.
\newblock \bibinfo{journal}{\emph{arXiv preprint arXiv:2310.18532}} (\bibinfo{year}{2023}).
\newblock


\bibitem[Muram et~al\mbox{.}(2018)]%
        {muram2018preventing}
\bibfield{author}{\bibinfo{person}{Faiz~UL Muram}, \bibinfo{person}{Barbara Gallina}, {and} \bibinfo{person}{Laura~G{\'o}mez Rodr{\'\i}guez}.} \bibinfo{year}{2018}\natexlab{}.
\newblock \showarticletitle{Preventing omission of key evidence fallacy in process-based argumentations}. In \bibinfo{booktitle}{\emph{2018 11th International Conference on the Quality of Information and Communications Technology (QUATIC)}}. IEEE, \bibinfo{pages}{65--73}.
\newblock


\bibitem[Murugesan et~al\mbox{.}(2023)]%
        {murugesan2023semantic}
\bibfield{author}{\bibinfo{person}{A. Murugesan}, \bibinfo{person}{I.~Hong Wong}, \bibinfo{person}{R. Stroud}, \bibinfo{person}{J. Arias}, \bibinfo{person}{E. Salazar}, \bibinfo{person}{G. Gupta}, \bibinfo{person}{R. Bloomfield}, \bibinfo{person}{S. Varadarajan}, {and} \bibinfo{person}{J. Rushby}.} \bibinfo{year}{2023}\natexlab{}.
\newblock \showarticletitle{Semantic Analysis of Assurance Cases using s (CASP)}. In \bibinfo{booktitle}{\emph{GDE Workshop in ICLP}}.
\newblock


\bibitem[Nair et~al\mbox{.}(2014)]%
        {nair2014extended}
\bibfield{author}{\bibinfo{person}{S. Nair}, \bibinfo{person}{J.~L. De~La~Vara}, \bibinfo{person}{M. Sabetzadeh}, {and} \bibinfo{person}{L. Briand}.} \bibinfo{year}{2014}\natexlab{}.
\newblock \showarticletitle{An extended systematic literature review on provision of evidence for safety certification}.
\newblock \bibinfo{journal}{\emph{IST}} \bibinfo{volume}{56}, \bibinfo{number}{7} (\bibinfo{year}{2014}), \bibinfo{pages}{689--717}.
\newblock


\bibitem[OpenAI(2023)]%
        {GPT4}
\bibfield{author}{\bibinfo{person}{OpenAI}.} \bibinfo{year}{2023}\natexlab{}.
\newblock \bibinfo{title}{GPT 4}.
\newblock
\newblock
\urldef\tempurl%
\url{https://openai.com/research/gpt-4}
\showURL{%
\tempurl}


\bibitem[{OpenAI}(2023)]%
        {OpenAI2023DevDay}
\bibfield{author}{\bibinfo{person}{{OpenAI}}.} \bibinfo{year}{2023}\natexlab{}.
\newblock \bibinfo{title}{New Models and Developer Products Announced at DevDay}.
\newblock \bibinfo{howpublished}{\url{https://openai.com/blog/new-models-and-developer-products-announced-at-devday}}.
\newblock
\newblock
\shownote{Accessed: 2024-01-14}.


\bibitem[Rushby(2013)]%
        {rushby2013logic}
\bibfield{author}{\bibinfo{person}{J. Rushby}.} \bibinfo{year}{2013}\natexlab{}.
\newblock \showarticletitle{Logic and epistemology in safety cases}. In \bibinfo{booktitle}{\emph{SafeComp}}. Springer, \bibinfo{pages}{1--7}.
\newblock


\bibitem[Rushby(2014)]%
        {rushby2014mechanized}
\bibfield{author}{\bibinfo{person}{J. Rushby}.} \bibinfo{year}{2014}\natexlab{}.
\newblock \showarticletitle{Mechanized support for assurance case argumentation}. In \bibinfo{booktitle}{\emph{New Frontiers in Artificial Intelligence: JSAI-isAI 2013 Workshops}}. Springer, \bibinfo{pages}{304--318}.
\newblock


\bibitem[SACM(2021)]%
        {SACM2021}
SACM \bibinfo{year}{2021}\natexlab{}.
\newblock \bibinfo{booktitle}{\emph{Structured Assurance Case Metamodel}}.
\newblock SACM.
\newblock


\bibitem[Sivakumar et~al\mbox{.}(2023)]%
        {sivakumar2023ArXiv}
\bibfield{author}{\bibinfo{person}{M. Sivakumar}, \bibinfo{person}{A. B.~Belle}, \bibinfo{person}{J. Shan}, {and} \bibinfo{person}{K. Khakzad~S.}} \bibinfo{year}{2023}\natexlab{}.
\newblock \showarticletitle{GPT-4 and Safety Case Generation: An Exploratory Analysis}.
\newblock \bibinfo{journal}{\emph{arXiv preprint arXiv:2312.05696}} (\bibinfo{year}{2023}).
\newblock


\bibitem[Viger et~al\mbox{.}(2023)]%
        {viger2023supporting}
\bibfield{author}{\bibinfo{person}{T. Viger}, \bibinfo{person}{L. Murphy}, \bibinfo{person}{S. Diemert}, \bibinfo{person}{C. Menghi}, \bibinfo{person}{A. Di}, {and} \bibinfo{person}{M. Chechik}.} \bibinfo{year}{2023}\natexlab{}.
\newblock \showarticletitle{Supporting Assurance Case Development Using Generative AI}. In \bibinfo{booktitle}{\emph{SAFECOMP 2023}}.
\newblock


\bibitem[Wei et~al\mbox{.}(2022)]%
        {wei2022chain}
\bibfield{author}{\bibinfo{person}{J. Wei}, \bibinfo{person}{X. Wang}, \bibinfo{person}{D. Schuurmans}, \bibinfo{person}{M. Bosma}, \bibinfo{person}{F. Xia}, \bibinfo{person}{E. Chi}, \bibinfo{person}{Quoc~V Le}, \bibinfo{person}{D. Zhou}, {et~al\mbox{.}}} \bibinfo{year}{2022}\natexlab{}.
\newblock \showarticletitle{Chain-of-thought prompting elicits reasoning in large language models}.
\newblock \bibinfo{journal}{\emph{NeuRIPS}}  \bibinfo{volume}{35} (\bibinfo{year}{2022}), \bibinfo{pages}{24824--24837}.
\newblock


\bibitem[Yuan et~al\mbox{.}(2016)]%
        {yuan2016automatically}
\bibfield{author}{\bibinfo{person}{T. Yuan}, \bibinfo{person}{S. Manandhar}, \bibinfo{person}{T. Kelly}, {and} \bibinfo{person}{S. Wells}.} \bibinfo{year}{2016}\natexlab{}.
\newblock \showarticletitle{Automatically detecting fallacies in system safety arguments}. In \bibinfo{booktitle}{\emph{PRIMA Workshops}}. Springer, \bibinfo{pages}{47--59}.
\newblock


\bibitem[Zhu et~al\mbox{.}(2023)]%
        {zhu2023large}
\bibfield{author}{\bibinfo{person}{Z. Zhu}, \bibinfo{person}{Y. Xue}, \bibinfo{person}{X. Chen}, \bibinfo{person}{D. Zhou}, \bibinfo{person}{J. Tang}, \bibinfo{person}{D. Schuurmans}, {and} \bibinfo{person}{H. Dai}.} \bibinfo{year}{2023}\natexlab{}.
\newblock \showarticletitle{Large Language Models can Learn Rules}.
\newblock \bibinfo{journal}{\emph{arXiv preprint arXiv:2310.07064}} (\bibinfo{year}{2023}).
\newblock


\end{thebibliography}

\appendix

\end{document}